# HIGHLY POROUS SUPERCONDUCTORS: SYNTHESIS, RESEARCH, AND PROSPECTS

**D. M. Gokhfeld[a, b], M. R. Koblischka[c], and A. Koblischka-Veneva[c]**
[a]*Kirensky Institute of Physics, Federal Research Center KSC SB RAS, Krasnoyarsk, Russia*
[b]*Siberian Federal University, Krasnoyarsk, Russia*
[c]*Shibura Institute of Technology,Tokyo, Japan*
e-mail: gokhfeld@iph.krasn.ru

This paper gives a review of studies of superconductors with a porosity above 50%. The pores in such superconducting materials provide refrigerant penetration, efficient heat dissipation and stable operation. Methods for the synthesis of the main groups of porous superconductors are described. The results of studies of the structural, magnetic, and electrical transport properties are presented, and the features of the current flow through porous superconductors of various types are considered.

## INTRODUCTION

Advances in raising the critical temperature and critical current open up new avenues to the widespread use of superconductors. Progress in the creation of superconducting high-current devices is based on theoretical studies of the pinning of Abrikosov vortices at artificial pinning centers [1]. Modern superconductingwires are capable of transmitting up to 500 A/cm per unit width at a temperature of 77 K [2], while massive superconductors can hold a magnetic field up to 18 T [3–5]. The development of new synthesis technologies and the creation of new forms of superconducting materials are promising areas of study, even if they do not lead to an increase in the critical temperature and critical current. The search for new forms has led to the advent of porous superconductors (PSs). The main feature of this form is its open porous structure.

Upon pulsed magnetization and high currents flowing in a superconductor, heating regions and hot spots may appear [6, 7]. If the outflow of heat is insufficient, then such heating regions spread, which leads to an increase in temperature and the destruction of the superconducting state in the entire volume. The uninterrupted functioning of superconducting devices requires constant removal of the generated heat. For effective cooling, it is necessary that liquid or gaseous refrigerant penetrate deep into the samples [8]. The access of the coolant deep into the material is possible if there are open macroscopic pores [9, 10] with a size of over 50 nm [11]. Thus, an important advantage of PSs is a large specific surface area, which provides the possibility of efficient cooling and preventing the growth of hot spots.

When creating a PS, scaling, i.e., the ability to create objects of various sizes, is available [12]. The synthesis of bulk superconducting samples larger than 5–10 cm is difficult due to cracking and uneven distribution of chemical elements. Another common problem for high-temperature superconductors (HTSCs), is the nonuniform oxygen saturation in the bulk. For materials with macroscopic pores, these problems are eliminated and PSs can have large sizes. $YBa_2Cu_3O_{7-\delta}$ (YBCO) porous materials with a foam structure were obtained for the first time in the early 2000s [13]. This discovery sparked a wave of works on the creation of new superconducting materials with macropores. A brief review of experimental works devoted to the study of porous HTSCs was presented in [12]. In the present review, we expand the range of PSs under consideration and analyze the distinctions and common features of their structural, magnetic, and electric transport characteristics.

## 1. TRAJECTORIES OF CURRENT IN POROUS SUPERCONDUCTORS

The thermal and mechanical properties of porous materials were analyzed using various structural models [14–23]. The shapes of structural elements (struts between pores) and the way they are packed determine the implements structure of the porous material [24]. Structural elements form a percolation system. In porous superconductors, superconducting clusters that ensure the flow of electric current, and clusters of pores, through which a coolant can flow, are formed. The coexistence of various types of percolation (polychromatic percolation [25]) is the property of PSs that is important for practical applications.

The trajectories of currents in a PS are more ramified and entangled than in dense bulk superconductors, and the current circulation pattern is more complicated [26–28]. PSs are internally heterogeneous at different scales. The ramified pattern of granules, pores, and clusters of porous materials can be characterized by the fractal dimension [11, 29, 30]. The fractal dimension of the pores, $D$, is related to the porosity $\varphi$ as $\varphi = 1 - (r_{min}/r_{max})^{3-D}$ [31], where $r_{min}$ and $r_{max}$ are the minimum and maximum radii of pores for which the dimensional analysis is carried out. The account for the fractal dimension of structural elements was made in a number of works [32–35] to describe the flow of fluids through porous media. The thermal and electrical conductivity of porous media also depends on the fractal dimension [36]. In nonhomogeneous superconductors, the fractal dimension of the boundaries separating the superconducting regions and pores affects the pinning, magnetic flux creep, and electrical transport properties [27, 37, 38].

The effect of the porosity of samples on their critical current was considered in [39–41]. Due to the presence of pores, the effective cross section for the current in a PS is smaller than in a superconductor without pores. With an increase in porosity, there is a sharp decrease in the area of contact areas of granules and, accordingly, a significant decrease in the critical current density through the PS.

When describing the magnetic and electric transport properties of PSs, the model of a two-level superconductor is used [42–48]. Inhomogeneous and polycrystalline superconductors are considered in the model as a Josephson medium [49–51] consisting of superconducting regions and weak links between them. The subsystem of weak links is responsible for the dissipation of energy in weak magnetic fields, while dissipation in strong magnetic fields occurs in the subsystem of superconducting regions. In magnetic fields $H$ smaller than the critical field of the Josephson medium, $H_{cJ}$, a polycrystalline superconductor behaves like a homogeneous superconductor, with a critical current density equal to the critical intergranular current density $J_{cJ}$. The magnetic flux begins to penetrate into the grain boundaries at $H > H_{cJ}$ [52]. In fields above ~0.01 T, the magnetic properties of the sample are mainly determined by the magnetization of the granules [53]. As a rule, the intragranular critical current density is several orders of magnitude higher than $J_{cJ}$ [54].

With increasing temperature or magnetic field, the critical values of both intragranular and intergranular currents decrease, but the corresponding functional dependences of the critical currents of these subsystems are different, since they are associated with different physical processes [48, 55]. The critical density $J_{cJ}$ is determined by the parameters of the grain boundary and the mutual orientation of anisotropic granules [44]. Also, the intergranular current is influenced by the pinning of the magnetic flux in the superconducting circuits around the pores [51, 56, 57]. Intragranular currents are controlled by pinning the magnetic flux at the inner and surface pinning centers [1]. Non-dissipative currents in a polycrystalline superconductor can circulate along two types of closed trajectories. A trajectory of the first type covers the entire sample, and the radius of current circulation $R_c$ is equal to the radius of the sample. A current with a density $J_{cJ}$ circulates along the perimeter of the sample through the granules and grain boundaries as long as the magnetic field is smaller than $H_{cJ}$. When $H > H_{cJ}$, the currents circulate inside the superconducting granules and the circulation scale is smaller than the sample size [55, 58, 59] corresponding to the average size of granules or clusters formed by several linked granules. Thus, the circulation radius $R_c$ is determined by both the material structure and external conditions.

To determine the average critical current density $J_c$ from magnetic measurements, the Bean formula $J_c(H) = \Delta M(H)/kR_c$ is used, where $\Delta M$ is the height of the hysteresis loop in A/m and the coefficient $k$ depends on the geometric shape of the sample; for polycrystalline samples, $k = 2/3$ is taken. The value of $\Delta M$ in an external magnetic field $H$ is determined as $\Delta M(H) = M{\downarrow}(H) - M{\uparrow}(H)$, where $M{\uparrow}(H)$ and $M{\downarrow}(H)$ are the magnetization values with increasing and decreasing $H$, respectively. For the correct determination of $J_c$, a correct choice of $R_c$ is necessary. For polycrystalline superconductors, the use of $R_c$ as the sample radius often leads to a significant underestimation of $J_c$.

A method for determining the current circulation scale was proposed in [59]. This method can be applied to superconducting single crystals and films with a small thickness $t$ such that $R_c/t > 2$. The demagnetizing factor of the samples that depends on the ratio $R_c/t$, influences the slope of the linear section of the dependence $M{\downarrow}(H)$ that appears immediately after switching the external field from increase to decrease. The circulation radius is determined from the equation $dM{\downarrow}(H)/dH = -\pi^2 R_c^3/[V\,(\ln(8R_c/t)-0.5)]$, here $V$ is the sample volume.

For samples with a negligible demagnetizing factor $R_c/t$ <<1) and polycrystalline superconductors, the following expression to determine the scale of current circulation was proposed in [60]:

$$R_c = \lambda(T)/[1 - (\Delta M(H_p)/2|M{\uparrow}(H_p)|)^{1/3}], \qquad (1)$$

where $\lambda$ is the magnetic field penetration depth for the superconductor at the measurement temperature and the values of $M{\uparrow}$ and $\Delta M$ are determined at $H = H_p$.

## 2. SYNTHESIS OF POROUS SUPERCONDUCTING MATERIALS

Superconducting materials with macropores can be obtained using various techniques. Previously synthesized PSs can be divided into several groups, differing in methods of preparation and structural features. To date, foams, porous polycrystals, sponges, and fabrics are known.

The synthesis of superconducting YBCO foams was carried out in two stages [13, 61]. At the first stage, $Y_2BaCuO_5$ (Y211) foam was prepared [62]. A standard technology for the production of ceramic foam was used, including impregnation of polyurethane foam with an aqueous suspension of Y211 powder and subsequent annealing. At the second stage, the top-seed growth (TSG) method was used. An $NdBa_2Cu_3O_{7-\delta}$ seed crystal was placed on the foam surface. The Y211 foam was impregnated with a melt of a mixture of barium cuprate and copper oxide [63], and the crystallization process into the YBCO phase repeating the structure of the initial foam occurred. The porosity of the YBCO foam obtained is determined by the structure of the polyurethane foam used in the first stage of the synthesis.

Porous polycrystals have a loose structure and contain a significant number of pores between crystallites, but the crystallites in these materials are the same as in bulk polycrystalline superconductors. $(Bi,Pb)_2Sr_2Ca_2Cu_3O_{10}$ (Bi2223) porous polycrystals were prepared by annealing a mixture of Bi2223 and $CaCO_3$ powders [64, 65]. The precursor was a polycrystalline superconductor of nominal composition $Bi_{1.8}Pb_{0.3}Sr_{1.9}Ca_2Cu_3O_{10}$ prepared from the corresponding oxides and carbonates $Bi_2O_3$, PbO, $SrCO_3$, and $CaCO_3$ by the standard solid phase synthesis technique [66]. From a mixture of milled Bi2223 and $CaCO_3$, tablets 20 mm in diameter and 4–5 mm in height are pressed. Highly porous samples are obtained by annealing tablets for 400 hours at a temperature of 820°C. The density of the samples obtained ranged from 1.55 to 2.26 g/cm$^3$, i.e., 26–38% of the theoretical density of Bi2223. The porosity of the prepared samples was controlled by the content of the $CaCO_3$ precursor.

Porous $MgB_2$ polycrystals were prepared by annealing a mixture of $MgB_2$ and Mg powders at a temperature of 900°C in a He atmosphere at a pressure of 1.5–1.7 bar [67, 68]. The samples obtained had a porosity of 30 to 84%. The porosity was controlled by the Mg content.

Porous YBCO polycrystals were prepared by annealing a mixture of YBCO particles and sugar [69, 70]. The porosity depends on the sugar content.

Superconducting sponges are formed by interwoven superconducting fibers. Sponges are synthesized based on various biopolymers and can repeat the structure of the biopolymers used. To synthesize superconducting sponges, a sol-gel process is used. YBCO sponges prepared by the annealing of a solidified mixture of polysaccharide dextran and an aqueous solution of yttrium, barium, and copper nitrates were studied [71–78]. Also, YBCO sponges were made using chitosan [77, 79], biogenic aragonite (cuttlefish bones) [80], graphene oxide [81], alginic acid [82], oligosaccharides [83], and xylan [84]. The creation of such sponges is possible on the basis of biopolymers replaced by Bi2223 [85]. The relative content of the polymer in the initial mixture seems to affect the porosity of the prepared superconducting sponge.

Nonwoven fabric is a felt-like material, similar to porous sponges, but formed by entangled superconducting nanofilaments. A nonwoven fabric of $Bi_2Sr_2CaCu_2O_8$ (Bi2212) nanowires was prepared by electrospinning (ES) [86, 87]. A solution of Bi, Sr, Ca, and Cu acetates, taken in a molar ratio Bi : Sr : Ca : Cu = 1 : 1 : 1 : 2 in propionic acid with the addition of polyvinylpyrrolidone to increase the viscosity was used. During electrospinning, nanowires were formed when an electric discharge was passed through a droplet of solution. The entangled nanofilaments formed a fabric. Finally, the fabric was held at 800°C in a pure $O_2$ atmosphere. The material has a highly porous structure, and the density of the fabric is 0.05 g/cm$^3$ that is only 0.72% of the theoretical density of Bi2212. A similar technology was used to synthesize fabrics from YBCO nanowires [88–90] and LSCO nanowires [91, 92]. Similar Bi2212 fabrics were obtained with an addition of Pb [87]. Doping with lithium makes it possible to lower the temperature of synthesis of Bi2212 fabrics [93]. Superconducting fabrics of YBCO [94, 95] and Bi2212 [96] with a similar structure were obtained by solution blow spinning (SBS). The fabrics obtained using ES or SBS are extremely loose; their porosity is up to 99.9%. Reducing the porosity of the fabrics is possible with additional processing of the material, e.g., by pressing and texturing.

## 3. ANALYSIS OF THE STRUCTURE OF POROUS SUPERCONDUCTORS

Figure 1 shows typical images of some PSs. YBCO foams have a cellular structure with large pores (Fig. 1a). In the central region of the sample in Fig. 1a, a remnant of the seed crystal is seen. Figures 1b and 1c show scanning electron microscopy (SEM) images of a Bi2223 porous polycrystal and a Bi2212 fabric, demonstrating the loose structure of these materials.

The mean pore sizes in the PSs under study (Table 1) were determined from SEM images of PSs with pores smaller than 100 μm and optical images of PSs with pores larger than 100 μm. The investigated materials can be divided into three groups differing in the mean pore size $d_p$: (1) large pores, $d_p$ ~ 1 mm (YBCO foams and polycrystals); (2) medium pores, $d_p$ ~ 1–100 μm (Bi2223 and $MgB_2$ porous polycrystals and sponges synthesized on the basis of biopolymers); and (3) small pores, $d_p$ ~ 1 μm (YBCO, Bi2212, and LSCO fabrics). Table 1 presents the mean crystallite sizes $d_s$ in the samples under study.

The crystal orientation and phase distribution in the PSs were investigated using electron backscattered diffraction (EBSD) [97]. EBSD is included in the set of functions of SEM devices and allows measuring crystal orientation with a relatively high spatial resolution. EBSD measurements of the cross sections of the YBCO foam struts [98, 99] show that almost the entire Y211 phase was converted to YBCO. Only tiny Y211 particles are distributed along characteristic faces. The orientation of the YBCO matrix in each strut corresponds to the geometric position of the strut in the foam. Thus, in contrast to the YBCO single crystal, the struts do not follow the dominant direction (0 0 1), but are oriented differently.

Due to charge effects that limit the spatial resolution, the standard EBSD technique is not suitable for studying the finest (~10–50 nm) Y211 particles in the YBCO phase. This problem is solved using the recently developed transmission EBSD technique that provides a resolution of up to 10 nm even in ceramic materials [100]. Owing to the transmission EBSD, it becomes possible to determine the orientation of crystallites in nanowires obtained using ES or SBS [101].

For Bi2223 porous polycrystals, the fractal dimension of the boundaries of pores and granules was determined from micrographs by covering with squares [27, 102]. The fractal dimension of boundaries on a plane can range from 1 (smooth boundaries and narrow pore size distribution) to 2 (maximum tortuosity and wide pore size distribution). The resulting fractal dimension of ≈1.8 indicates a strong tortuosity of the pore boundaries and the presence of both very small and very large pores.

a

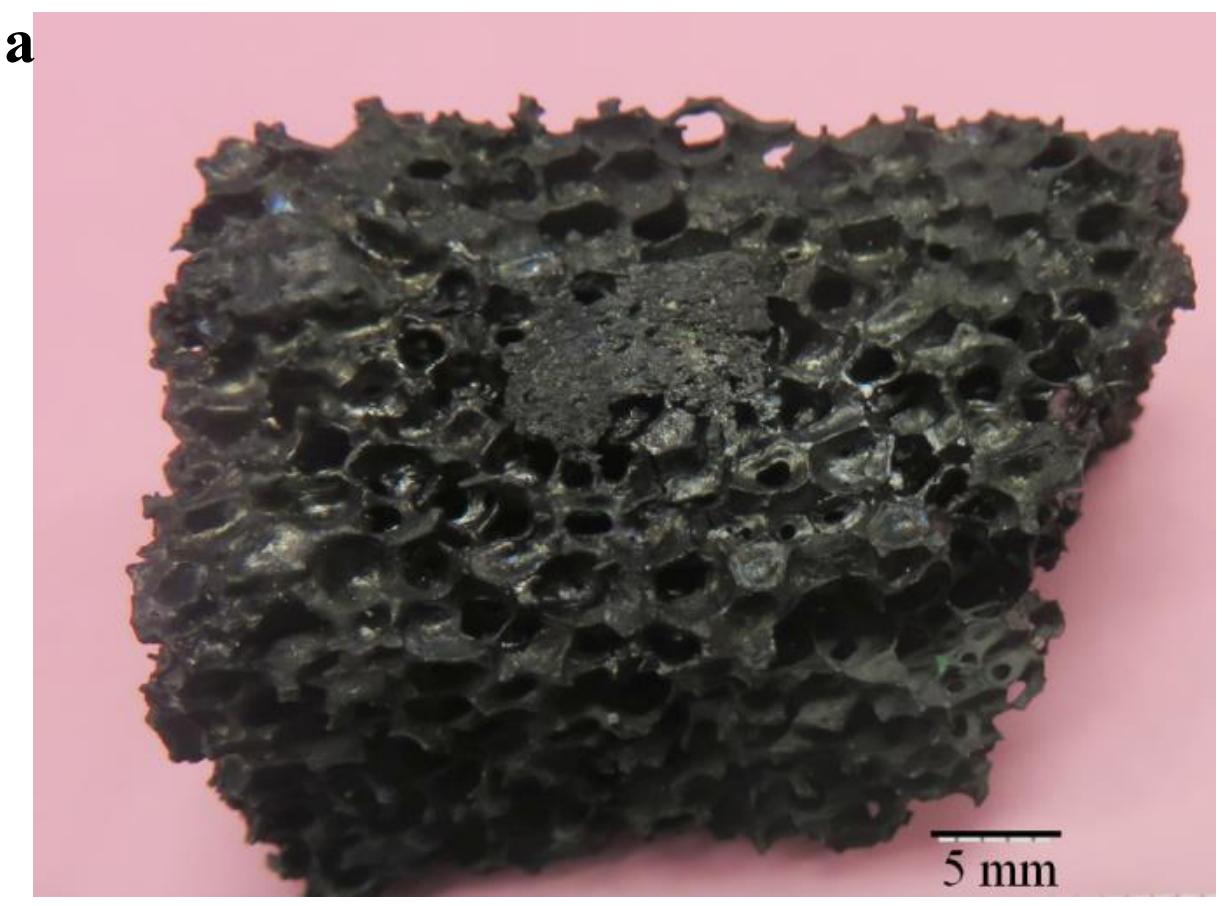

b

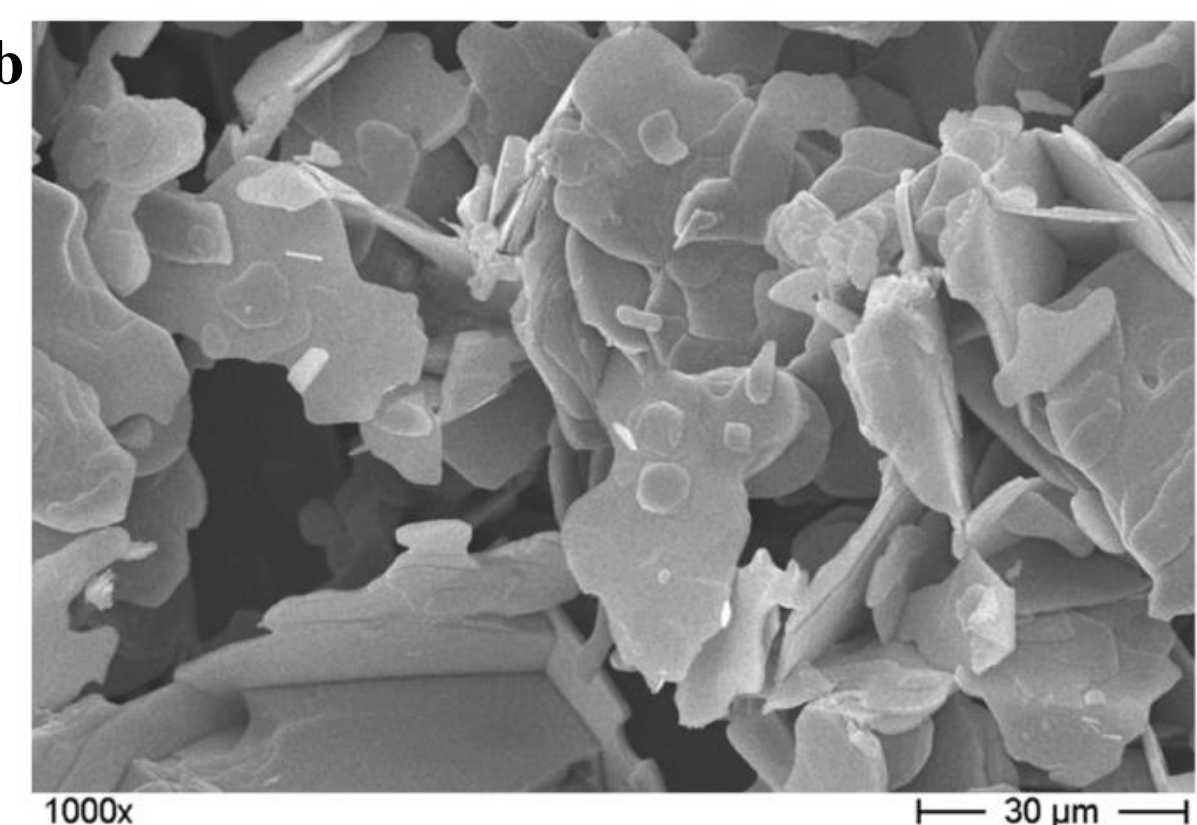

c

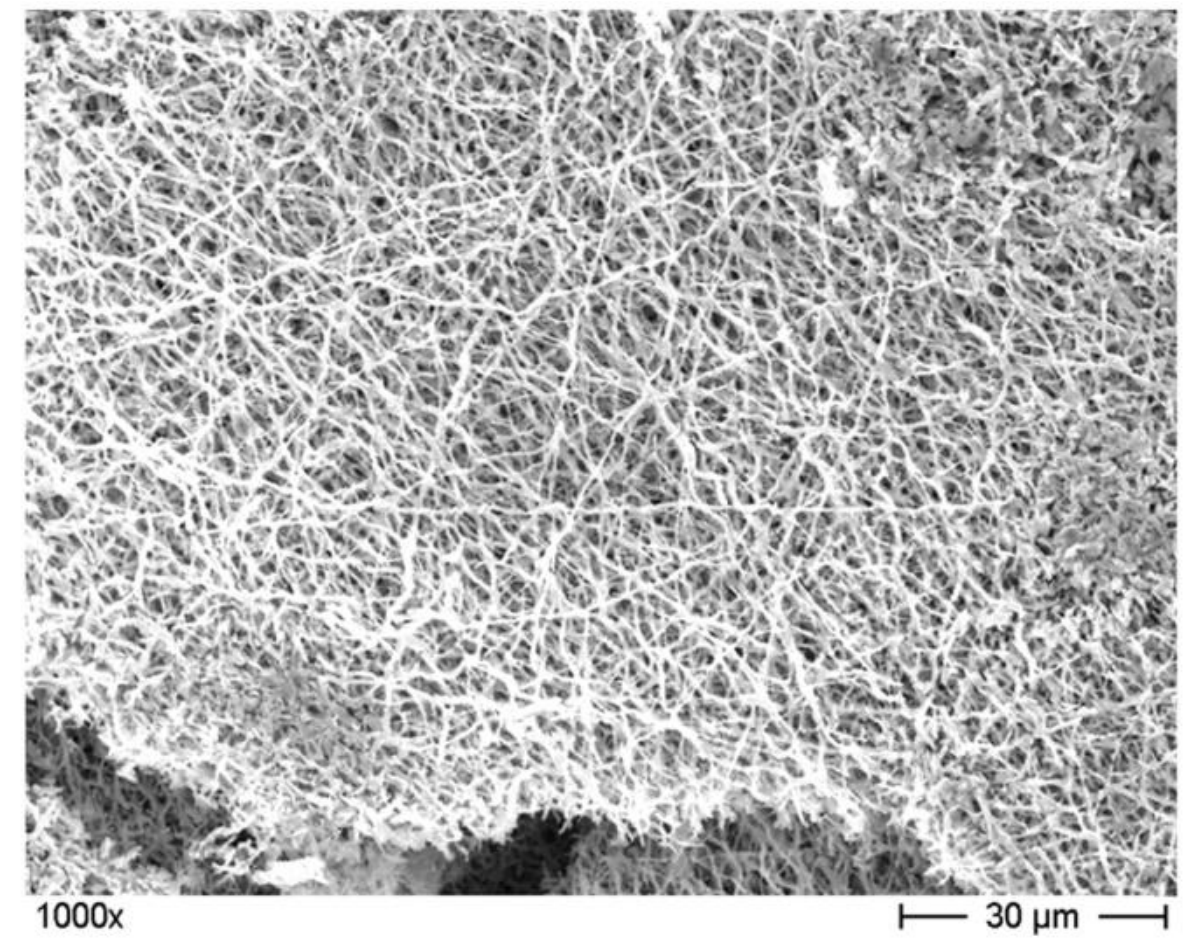

**Fig. 1.** (a) Photograph of YBCO foam, (b) microphotographs of Bi2223 porous polycrystal, and (c) Bi2212 fabric.

It should be noted that the structure of PSs has not been studied in sufficient detail in comparison with other porous materials [8, 11]. New information on the features of the PS structure and the formation of an end-to-end pore cluster can be obtained using X-ray computed tomography [103], nuclear magnetic resonance, mercury porosimetry, and capillary flow porosimetry [104, 105].

**Table 1.** Structure parameters (maximum porosity, mean pore size $d_p$, and crystallite size $d_s$), magnetic hysteresis height $\Delta M$ at $H = 0$ and critical temperature $T_c$

| Material | | Synthesis | Porosity, % | $d_p$, μm | $d_s$, μm | $\Delta M$, A m/kg | $T_c$, K |
|---|---|---|---|---|---|---|---|
| YBCO foam | | TSG,Y211 foam [13] | 70-80 | 300-1300 | 100 | 20* | 91 |
| Polycr. | YBCO | YBCO + sugar [69,70] | 60 | 100-1000 | 100-800 | - | 93 |
| | Bi2223 | Bi2223 + $CaCO_3$ [64] | 74 | 1-10 | 10 | 50**, 2* | 110 |
| | $MgB_2$ | $MgB_2$ + Mg [67] | 84 | 1-10 | 1-10 | 10*** | 38 |
| YBCO sponge | | Sol-gel [71,79] | - | 1-100 | 0.1-1 | 20**, 1* | 75-93 |
| Fabric | YBCO | ES [88,89], SBS [94] | - | 1-2 | 0.2-0.6 | - | 92-93 |
| | Bi2212 | ES [86,87], SBS [96] | 99.9 | 1-2 | 0.32 | 2** | 75-84 |
| | LSCO | ES [91,92] | - | 1 | 0.12 | - | 19 |

*At $T = 77$ K. **At $T = 10$ K. ***At $T = 20$ K.

## 4. MAGNETIZATION AND CURRENTS IN POROUS SUPERCONDUCTORS

Does high porosity affect the magnetic and electrical transport properties of superconductors? The features of magnetic flux pinning have been studied in YBCO foams [28, 39, 99, 106–111], Bi2212 fabrics [87, 93, 109, 112–115], and porous polycrystals of Bi2223 [64, 116–120] and $MgB_2$ [67]. For YBCO fabrics, porous polycrystalline YBCO, and LSCO fabrics, the results of magnetic measurements have not been presented in published articles. However, it can be assumed that the magnetic characteristics considered here are characteristic of all PSs.

For most of the created highly porous superconductors, magnetization hysteresis loops have been measured at different temperatures. The magnetic hysteresis of porous superconductors (Fig. 2a) has the same features as the typical hysteresis loops of polycrystalline superconductors. These features include the increasing asymmetry of the loop with respect to the axis $M = 0$ with increasing temperature. This asymmetry is caused by the contribution from the equilibrium magnetization of the granule surfaces [44, 121, 122]. Another feature is the presence of an irreversibility field $H_{irr}$; at $H = H_{irr}$, the branches $M\uparrow(H)$ and $M\downarrow(H)$ of the loop merge. For the magnetic hysteresis of Bi2223 porous polycrystal shown in Fig. 2a [117, 120], the irreversibility field is 5 T at $T = 40$ K. With increasing $T$, $H_{irr}$ decreases; at $T = 80$ K, the dependence $M(H)$ is irreversible only in the range of fields from –0.15 to 0.15 T.

The observed asymmetry of the magnetization hysteresis loop makes it possible to determine the current circulation scale. Using formula (1) for magnetic hysteresis at $T = 4.2$ K (Fig. 2a), we estimate $\lambda/R_c \approx 0.03$. The magnetic field penetration depth $\lambda$ for Bi2223 is 150 nm [54, 123]; hence, $R_c \sim 5$ μm. This scale corresponds to the size of Bi2223 crystallites in the *ab* plane. The value of $R_c$ obtained confirms the fact that, in strong magnetic fields, the circulation of currents occurs in the *ab* planes of crystallites.

The total magnetization of a polycrystalline sample along the external field is the projection of the magnetic moments created by circulating currents in all crystallites. The proportionality between $\Delta M$ and the circulating current allows one to determine the critical current density from the magnetic hysteresis using the Bean formula. In this case, to determine the values of $\Delta M$ in A/m, one should use the physical density of the crystallites rather than the density of the porous material.

For Bi2223 porous polycrystal and Bi2212 fabric, the critical current density, determined from the magnetization hysteresis loops, decreases with increasing magnetic field as $J_c(H) \sim H^{-\alpha}$ (Fig. 2b). Using the modified dependence $J_c(H)$ [122], the coefficient $\alpha \approx 0.6$ was obtained for the PSs Bi2223 and Bi2212 at all temperatures [112, 113]. The dependences of the pinning force $F_p$ on the magnetic field, determined by the formula $F_p(H) = \mu_0 H J_c(H)$, have a maximum at $H \approx 0.11–0.13\ H_{irr}(T)$ (Fig. 2c). The critical current density $J_c$ at $H = 0$ decreases exponentially with increasing temperature. The exponential decreasing dependence $J_c(T)$ is typical of the superconductors Bi2212 and Bi2223 [124, 125]. This temperature dependence of $J_c$ corresponds to the collective pinning of the vortex lattice at weak pinning centers [126]. Due to thermal fluctuations, the vortices detach from the pinning centers, which leads to a creep of the vortex lattice and the rise of dissipation.

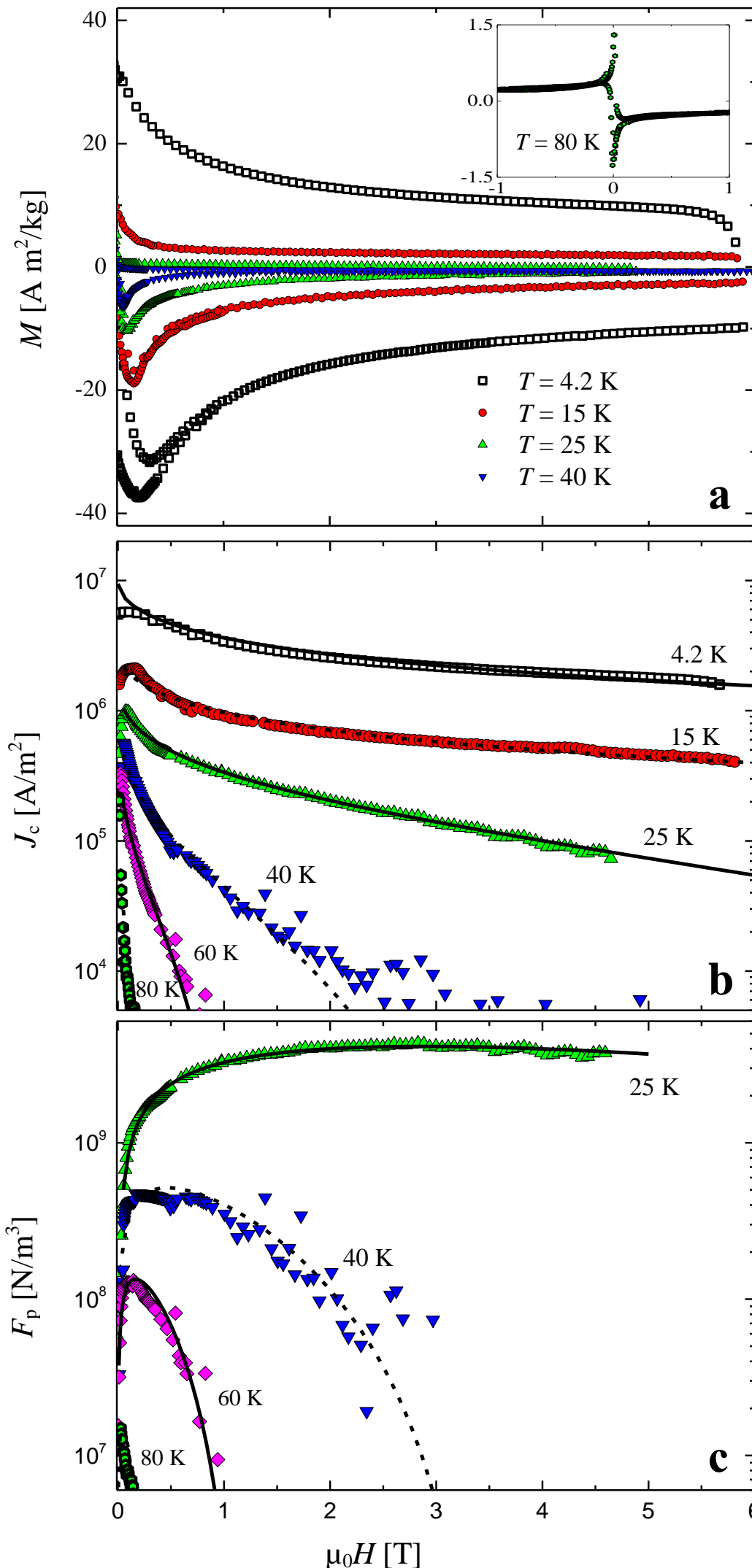

**Fig. 2.** (a) Magnetization hysteresis loops and field dependences of (b) the critical current density and (c) pinning force of porous Bi2223. The curves represent modified dependences of the critical current density [112, 122].

Measurements of the field trapped in YBCO foam demonstrate the presence of two current subsystems. Figure 3a shows the distribution of the trapped field $B(x, y)$ over a $\approx 50 \times 20 \times 20$ mm foam sample upon cooling in a magnetic field of 0.5 T [28]. The presented distribution has a main peak and several small peaks. The main peak of the trapped field is caused by the currents flowing along the perimeter of the sample. Numerous smaller peaks are created by currents circulating in superconducting loops around individual pore clusters. The position of the peaks is reproduced on repeated scanning, although their height may change. It should be noted that only the $z$ component of the internal field generated by the circulating currents was determined in the measurements. However, currents of both types (with different circulation scales) flow through the superconducting regions along complicated three-dimensional trajectories that do not always coincide with the sample plane. Therefore, only the $(x, y)$-projection of real circulating currents is reflected in the measurements of the trapped field.

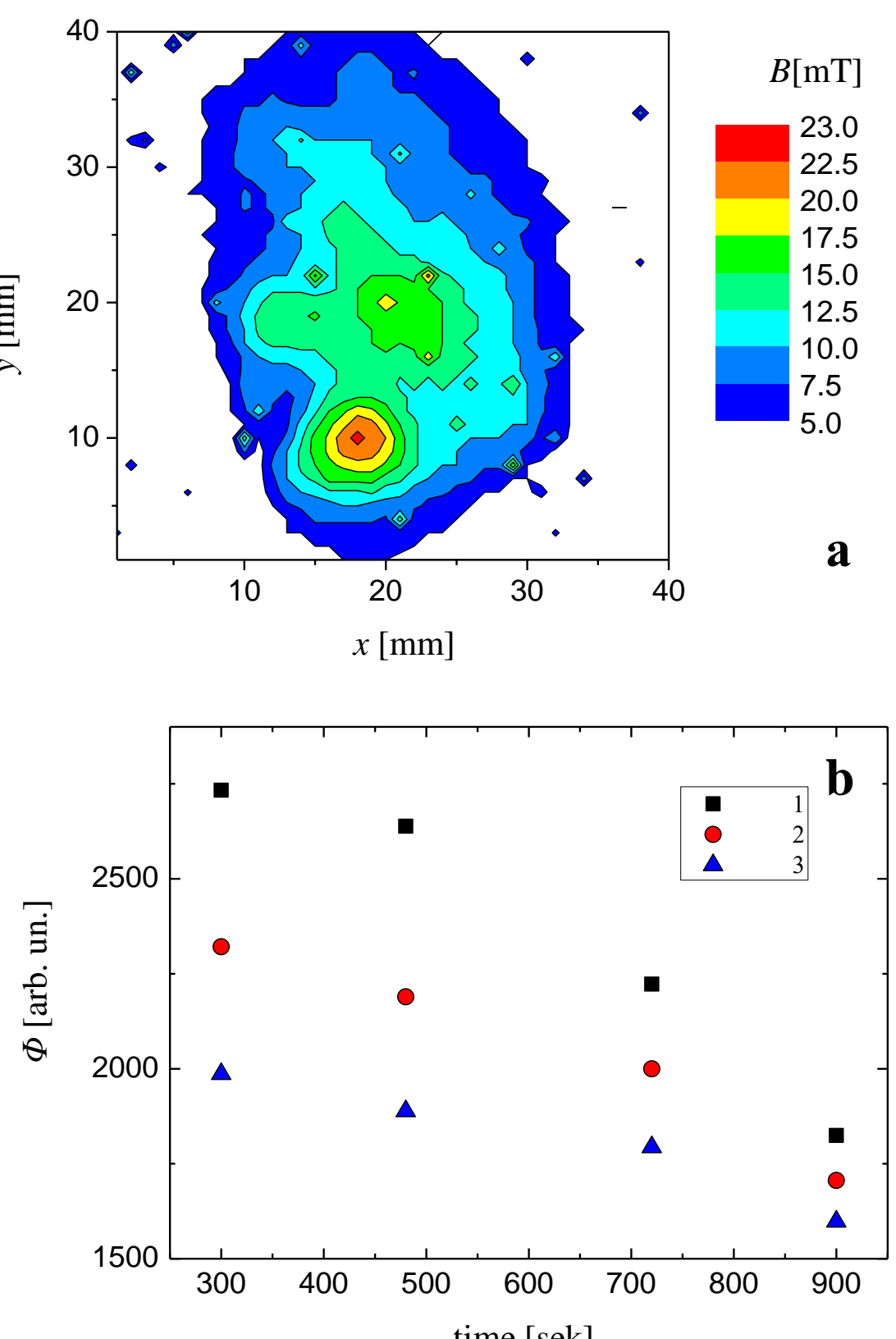

**Fig. 3.** (a) Trapped field $B$ in YBCO foam upon cooling in a magnetic field of 0.5 T and (b) time dependence of the magnetic flux $\Phi$ around some peaks.

Measurements of the magnetic flux relaxation reveal interesting features of magnetic flux pinning and creep in YBCO foam. In measurements on individual struts, when all effects from pores are absent, the relaxation coefficient of the trapped magnetic flux is about 4% in the temperature range 20–60 K and magnetic fields up to 2 T [111]. This indicates the realization of strong pinning provided by small Y211 particles distributed in the YBCO matrix. Due to strong pinning, the values of $H$irr for YBCO struts are higher than in Bi2212 and Bi2223 porous polycrystals and the position of the maximum of the dependences $Fp(H)$ in the temperature range 60–85 K is in the field $H \approx 0.5\ H_{irr}(T)$ [99, 108, 110].

The relaxation of the trapped magnetic flux in large lumps of foam has contributions from all types of currents: the currents flowing along the perimeter of the sample and responsible for the main peak; the currents around the pores that are responsible for narrow secondary peaks; and the local currents inside the struts. The total relaxation factor of the magnetic flux is 8% at $T$ = 77 K [111]. Upon the relaxation of the magnetic flux, a rearrangement of the circulating currents occurs. Due to the rearrangement of currents, the relaxation rate of the secondary peaks of the trapped field begins to increase after a certain characteristic time (Fig. 3b). The relaxation coefficient for secondary peaks reaches 20%.

The current–voltage characteristics of Bi2223 porous polycrystals [27, 102, 127, 128] and Bi2212 fabrics [115] have a nonlinear form, typical of bulk HTSCs [129]. With increasing current, the voltage drop across the samples gradually increases. The critical current transport density $Jc$ determined at $T$ = 77 K from the parameter 1 μV/cm is ~1000 A/cm$^2$ for the YBCO foam [13] and up to 10 A/cm$^2$ for the Bi2223 porous polycrystal [128].

At $H > H_{cJ}$, in polycrystalline superconductors, nonzero resistance appears. Since the area of contact points of neighboring granules is one to two orders of magnitude smaller than that in dense polycrystalline HTSCs, the resistance of Bi2223 porous polycrystals is more sensitive to changes in the external magnetic field [102, 118, 130]. The magnetoresistance of the Bi2223 porous polycrystal exhibits an inverse hysteresis [130]: the resistance with an increase in the external field is greater than that with a decrease. This behavior is caused by the trapping and compression of the magnetic flux in the grain boundaries of polycrystalline superconductors [131].

The temperature dependences of the resistance of polycrystalline superconductors have a characteristic two-step form: a jump in resistance at the critical temperature $T_c$ and an extended decrease in resistance to 0 as $T$ decreases from $T_c$ to $T_{c0}$. A similar picture was observed for YBCO foam [106], YBCO porous polycrystals [70], Bi2223 [64, 128], and Bi2212 fabrics [115]. The observed behavior of the resistance corresponds to the two-level model of a polycrystalline superconductor. At $T = T_c$, superconducting granules pass into the superconducting state, which leads to a sharp jump in resistance. In the temperature range $T_{c0} < T < T_c$, the resistance decreases to 0, due to the establishment of superconductivity in the Josephson network formed by granules and grain boundaries. The width of the resistive transition in a zero magnetic field, $\Delta T_c = T_c - T_{c0}$, is equal to ≈3 K for YBCO foam and YBCO polycrystal and Bi2223 [12]. In Bi2212 fabrics, the resistive transition is more extended, $\Delta Tc$ ~ 20–50 K [12].

To compare different PS samples, it is convenient, in our opinion, to use a parameter such as the height of the magnetic hysteresis $\Delta M$ in zero magnetic field at a low temperature (10 K) or at a technologically important boiling point of liquid nitrogen (77 K). Such values of $\Delta M$ can be estimated for a number of samples from the graphs and data presented in published articles (Table 1). Of course, for a full comparison of the properties of various PSs, it is necessary to fill all the empty cells in Table 1. Nevertheless, the available data allow us to draw some conclusions about the effect of structural features on the magnetic properties. For all PSs, a correlation between the values of $\Delta M$ and $d_s$ is observed. This correlation indicates that the magnetization is mainly determined by the circulation of currents on the scale $2R_c \approx d_s$. The values of the local intragranular current density for various HTSCs and $MgB_2$, apparently, differ insignificantly [115, 143]. The highest values are found in YBCO foam, for which $\Delta M$ (77 K) ≥ 20 A m/kg. First of all, the high values of $\Delta M$ are associated with the large size of the superconducting struts in the foam (~0.1 mm), which determines the scale of current circulation in strong fields. Also, the value of $\Delta M$ is positively influenced by the low anisotropy of YBCO and strong pinning in the struts, which lead to high values of the current density.

Experimental studies show that high porosity does not lead to significant differences in the behavior of magnetization and resistance in comparison with the temperature and field dependences of the magnetization and resistance of dense polycrystalline superconductors.

## 5. PROSPECTS FOR USE OF POROUS SUPERCONDUCTORS

YBCO foam is currently the most suitable material in terms of characteristics important for practical applications. The foam exhibits the best mechanical properties and the highest $Jc$ values among the materials studied. Polycrystalline PSs, fabrics, and sponges also have potential for development and applications.

Polycrystalline PSs can be used as a basis for the creation of textured superconducting materials [132, 133]. An advantage of porous polycrystals, which is important for many researchers, is the relative simplicity of their preparation. The polycrystalline nature is the main weakness of all types of PSs obtained to date. Measures for enhancing intercrystalline connectivity and a decrease in the disordering of crystallites [134] can significantly improve the transport critical current density [54].

The mechanical properties of the porous materials presented are inferior to those of dense superconductors. An improvement of both mechanical and current-carrying properties in PSs can be achieved by creating composites with silver [119, 135, 136]. To improve the mechanical properties, it was proposed in [13] to impregnate a porous material with rubber. However, such an impregnation will prevent the penetration of he coolant into the interior of the sample and eliminate the important advantage of the porous structure. Acceptable methods of reinforcement should not prevent refrigerant from penetrating deep into the sample. The use of a steel frame to reinforce a PS satisfies this condition and will allow it to withstand significant trapped fields.

PSs can find their niche among materials for a variety of applications that require both large size, light weight, and high cooling rates. Due to the presence of pore clusters, a liquid or gaseous refrigerant such as liquid nitrogen can be pumped through the PS and provide extremely efficient cooling. This prevents the growth of hot spots arising from the flow of currents close to critical. An efficient heat removal makes PSs attractive for use as short-circuit current limiters [137, 138].

Due to their low physical density, PSs are suitable for creating ultra-light devices [139]. Particularly stringent requirements to the mass of working elements are imposed in aviation and space technology. There are projects for the use of superconductors for docking systems of spacecraft [140, 141], devices for micrometeorite protection and space debris collection [142, 143], and aircraft electric motors [5, 144, 145]. The use of PSs in these devices should provide additional advantages in ensuring stable operation and reducing weight. For some problems, e.g., to increase the dynamic stability during magnetic levitation [146], a combination of porous and dense superconductors may be a good solution.

We hope that this review of works on porous superconductors will give impetus to further research on these promising materials. It is expected that new results will be obtained through the development of methods of synthesis, computer simulation, and machine learning.